\documentclass[reprint,aps,prb,amsmath,amssymb]{revtex4}

\usepackage[T1]{fontenc}
\usepackage[]{graphicx}
\usepackage{times}
\usepackage{bm}
\usepackage{color}

\newcommand{\eq}{\begin{eqnarray}}
\newcommand{\en}{\end{eqnarray}}

\newcommand{\ud}[1]{{#1^{\dagger}}}
\newcommand{\be}{\begin{equation}}
\newcommand{\ee}{\end{equation}}
\newcommand{\bea}{\begin{eqnarray}}
\newcommand{\eea}{\end{eqnarray}}

\newcommand{\ket}[1]{\ensuremath{| #1 \rangle}}

\def\bra#1{\mathinner{\langle{#1}|}}
\def\ket#1{\mathinner{|{#1}\rangle}}
\newcommand{\braket}[2]{\langle #1|#2\rangle}

\begin{document}

\title{Influence of the Pauli exclusion principle
on  scattering properties of cobosons}

\author{A. Thilagam}
\email{thilaphys@gmail.com}
\affiliation{\rm Information Technology, Engineering and Environment, \\
University of South Australia, \\ South Australia  5095 \\ 
Australia }


\begin{abstract}
We examine the influence of the Pauli exclusion principle
on the scattering properties of composite bosons (cobosons) made  of two fermions, such as 
the exciton quasiparticle. 
The scattering process incorporates boson-phonon interactions that arise due to lattice vibrations.
Composite boson scattering rates increase with the entanglement 
between the two fermionic constituents, which comes with a
 larger number of available single-fermion states. 
An important role is  played by probabilities associated with
accommodating an incoming  boson
among the remaining  unoccupied Schmidt
modes in the initial composite system.
While due attention is given to  bi-fermion bosons, 
the methodology is applicable to any composite boson
 made up of smaller  boson fragments.  
Due to  super-bunching in a  system of multiple boson condensates
such as bi-bosons, there is enhanced scattering  associated with  bosons occupying
 macroscopically occupied Schmidt modes, in contrast to the system of bi-fermion pairs.
\end{abstract}

\maketitle

\section{Introduction}

Composite bosons   \cite{compra1,compra2,comrep,Combescot2011,romb,Law,woot} 
that fall within the spectrum bounded by ideal  bosons and fermions
have been the subject of many recent works
 \cite{avancini,Sancho,Rama1,Rama2,brou,Gavrilik,Gav2,Tichy12,thilzil,thilchem}.
While several bosons may occupy the same
state, multiple occupation is inhibited in the case of  two fermions,
 due to the Pauli exclusion principle.  The difference between bosons and fermions
is reflected in all  basic and experimental
studies due to the  symmetrization postulate, and 
interferences that arise through the superposition principle. 
For composite boson made of 
an even number of fermions, also known as ``cobosons" \cite{comrep,Combescot2011}, 
the Pauli principle does not influence
the dynamics of the two highly  entangled fermions.
In this case, the constituent fermions seldom compete for single-particle states.
The Pauli principle, although still omnipresent, therefore
  does not influence composite bosons with low occupation probabilities.
A range of phase-space filling  effects and  
commutation relations 
arise due to the emergence and pronounced
governance  of the Pauli principle beyond a critical
level of occupation probabilities of the 
constituents of the coboson species.

Recent studies on composite bosons made of two distinguishable entangled constituents
such as the two-fermion boson system,
have shown the subtle links between entanglement and indistinguishability,
through the diminishing effects of the  Pauli exclusion principle  with 
increase in entanglement 
  \cite{Combescot2011,Law,woot,avancini,Rama1,Rama2,tichyprl}.
The term ``entanglement" refers to the situation in which
 individual non-interacting constituents of
a quantum system are influenced by one another,  with a collective
wavefunction describing the quantum properties  of the system.
An algebraic description of composite bosons
 from the perspective of quantum information \cite{Combescot2011,Law,woot}
provides  insight  to
the microscopic quantum description of many body systems.
The purity $P$ of the single-particle density matrix is a quantitative
indicator for entanglement of a system of  constituent 
fermions  \cite{Combescot2011,Law,woot}. 
Deviations from unity of the  ratio, $\alpha_{N+1}$=$\sqrt{\chi_{N+1}/\chi_N}$,
to be defined below, where $\chi_N$ is the normalization term associated with    $N$ cobosons,
provides a measure of ``compositeness"  of systems of boson and fermion 
constituents \cite{Combescot2011,Law,woot}.
Composite bosons with minimal deviations can be approximated
as ideal bosons.  The upper and lower bounds
to $\chi_N$ in terms of the purity $P$ of the single-fermion reduced state,
show convergence at  small purities \cite{Combescot2011,Law,woot,Rama1,Rama2,Tichy12}. 
At higher purities, the  bounds  become inefficient \cite{Tichy12,tichyprl,tichy2013a,tichy2013b}
as factors other than $P$ may control the behavior of the 
composite bosons. Tighter bounds for the normalisation factor $\chi_N$ and for
 the normalisation ratio $\chi_{N+1}/\chi_N$ for two-fermion cobosons were recently obtained in terms of the
 purity $P$  and the largest eigenvalue $\lambda_1$ of 
the single-fermion density matrix \cite{tichy2013b}. 
Due to incorporation of more information through $P$ and $\lambda_1$,
the improved results \cite{tichy2013b}
enabled convenient  evaluation of the  normalisation factor at large composite numbers $N$.

 In our earlier works \cite{thil1,thil2a,thil2b}, 
the  composite nature of excitons was neglected, partly due to the 
simplicity and effectiveness of the ideal boson
description  of the exciton system at low densities \cite{hana,taka}.
When the mean
inter-excitonic distance greatly exceeds the exciton
Bohr radius,  the correlated electron-hole quasi-particle  can be
considered  structureless. The assumption of the   spin independent 
exciton model breaks down when the dynamics of interacting excitons  
is influenced by  the Pauli exclsuion principle.
Further neglect of  Pauli exclusion as the inter-exciton separation
is decreased, will result in increased   non-Hermitian features which may distort  computed 
exciton lifetimes. Combescot and coworkers have proposed a ``commutator formalism" \cite{comrep}
to incorporate the inter-excitonic Pauli exclusion scatterings which are critical
to explaining optical features not associated with coulombic interactions between fermions.

The case of the high-density electron-hole system with 
excitonic instability has also been studied using techniques based on
the generalized random-phase approximation \cite{ina}, and  the
 vertex-equation extension \cite{chuchang} of the 
Bardeen, Cooper, and Schrieffer (BCS) theory \cite{bar1,bar2}.
In a recent work,  Koinov \cite{koi} employed the BCS and Bethe-Salpeter equations
to highlight the appearance of a  secondary peak in the optical
spectrum that can be  linked to an excitonic  phase of high density. 
Imamoglu \cite{imag} examined the limitations imposed by
Pauli exclusion of   fermions in  exciton-phonon interactions, and obtained
results showing a dependence of scattering times  on the density of the composite fermionic 
species.  In this work, we examine the influence of 
the Pauli exclusion principle during scattering 
of the bi-fermion excitons by phonons which arise from lattice vibrations.
We focus on the entanglement attributes of the scattered composite boson
system, thereby extending the earlier work of Imamoglu \cite{imag},
to include quantum information theoretic  factors such as  purity, $P$, 
and  the  normalization ratio of composite-boson states.
This approach will provide a realistic assessment  of the 
 Pauli exclusion effects on the lifetimes of the scattered excitons
at high  densities of   correlated
electron-hole pair systems.

The results of this work will  also be of interest
to composite boson systems  that are made of two distinguishable
bound bosonic constituents, otherwise known as bi-boson composites \cite{tichy2013a}.
Based on the interplay of  interactions between  boson constituents
and the global  composite,  bi-bosons may  operate in the
 super-bosonic phase in which
the boson constituents display enhanced bunching \cite{tichy2013a}.
A bunching process is associated with the  tendency for particles
 to be distributed in preferred collective modes instead of a random
Poissonian type distribution. In super-bunching, a specific mode
for boson occupation is preferred at the expense of other modes.
As the number of composite boson is increased,  a single mode occupied by a
boson attracts  further occupation  which results in macroscopic occupation of bosons
in the preferred  mode \cite{tichy2013a}. The super-bunching behavior therefore reduces  the occupation of bosons 
present in other modes.

There   results obtained for bi-bosons may be applied 
to  complex aggregate systems containing several electron-hole pairs. 
In a recent work \cite{dropt}, electron-hole aggregates were seen to 
give rise to a new form of stable  quasiparticle states known as quantum droplets.
A correlated electron-pair aggregate of large size (ten times
the size of a single exciton) in 
GaAs \cite{dropt} was observed using experimental techniques.
The minimum requirement of
four  electron-hole pairs for stability is novel
as the electrons and holes exist in  unpaired configurations,
yet the  quantum droplet  appear  as a collective
boson entity.

This  paper is organized as follows. In Section \ref{cobo} we provide 
a brief review of the  physics of cobosons,  and  examine the 
characteristics of  the lower and upper limits to the 
normalisation ratio  in composite boson systems.
In Section \ref{numb}, we discuss the subtle difference between the  electron-hole pair numbers
and the boson number, and  provide a
 physical interpretation of the number-operator for composite bosons.
We also examine the conditions under which an orthogonal
fermionic fragment state is formed when a coboson 
 dissociates into constituents in 
 orthogonal subspaces. In Section \ref{fluc},
we derive expressions related to the fluctuation to the mean number of 
 correlated coboson constituents. In Section \ref{pho}, we examine the 
BCS variational ansatz  in the context of excitonic systems, 
and establish the  links between the BCS state parameters,
 purity $P$ and the normalization ratio $\alpha_{N+1}$. Using the results in Section \ref{ansatz},
we obtain the scattering rate of  composite exciton condensates due to lattice vibrations
in Section \ref{phos}, with our main result
showing the dependence of this rate on 
the normalization  ratio. 
In Section \ref{sub}, the composite boson made of 
two bound bosonic constituents or bi-boson systems is examined qualitatively
in the context of the findings in Section \ref{pho}.
We present our conclusion in Section \ref{con}

\section {Cobosons  states : Preliminaries}\label{cobo}

The  creation operator of a coboson made of distinguishable fermions can be written in the Schmidt 
decomposition as \cite{Combescot2011,Law,woot}
\be
 \label{boper}
\hat c^\dagger=\sum_{j=1}^S \sqrt{\lambda_{j}}\; \hat a_{j}^\dagger \hat b_{j}^\dagger
=: \sum_{j=1}^S \sqrt{\lambda_j}\;\hat d^\dagger_j, 
\ee
  where  $\lambda_j$ are the Schmidt coefficients,  $ \hat a^\dagger_{j}$ and 
$ b^\dagger_{j}$ are fermion creation operators associated with each Schmidt mode, and
$S$ denotes the total number of Schmidt coefficients \cite{schmi}.
The operator $\hat d_j^\dagger$ creates a bi-fermion product state in the mode $j$,
hence the operator $\hat c^\dagger$ appears as
 a weighted superposition of all bi-fermion operators that are
distributed among the Schmidt
modes for the two constituents operators, $ \hat a^\dagger_{j}$ and 
$ b^\dagger_{j}$.  The distribution of $\lambda_j$ = $\vec \Lambda=(\lambda_1, \dots , \lambda_S)$
( $\lambda_1 \ge \lambda_2 \ge \dots  \ge 0$) fulfills  $\sum_{j=1}^S \lambda_j =1$.
The purity $P=\sum_{j=1}^S \lambda_j^2$ is related to the Schmidt number $K$ \cite{grobe} via $K=1/P$, where 
the latter quantifies the correlations between the fermions. 
 In the case of the exciton, a  large ${\cal K}$ implies a highly correlated  electron-hole
pair linked to high binding energies. A less tightly bound exciton is linked to
 a more  distinguishable (and less entangled) electron and hole system.

The operators, $\hat c$ and $\hat c^\dagger$ obey the approximately bosonic commutation relations
\bea
[\hat c ,\hat c] &=& [\hat c^\dagger ,\hat c^\dagger] = 0 , \nonumber \\ {} \label{ccdagger}
[\hat c ,\hat c^\dagger] &=&  1 + t \sum_{k=1}^S \lambda_k (\hat a_{k}^\dagger \hat a_{k} + b_{k}^\dagger \hat b_{k}),
\eea
with $t=1$ ($t=-1$) for bi-bosons (bi-fermions). This results in 
differences between cobosons, depending on  their constituents (bosons or fermions).

 The state of $N$ composite bosons  can be expressed as
a superposition of $N$ bi-fermions or $N$ bi-bosons 
 as follows \cite{Law,Tichy12,tichy2013a}
\bea 
\label{istateN}
\ket{N} = \frac{1}{\sqrt{ N! \chi_N^{J}} } \ket{\psi_N}
= \frac{1}{\sqrt{ N! \chi_N^{J}}}
 \left(  \hat c^\dagger \right)^N \ket{0}
\eea 
where the normalization factor is given by 
$\chi_N^{J}$=$\chi_N^{B}$ ($\chi_N^{F}$)
in the case of bi-bosons (bi-fermions).
The states $\ket{\psi_N} = (\hat c^\dagger)^N\ket{0}$ are  not normalized as $\braket{\psi_N}{\psi_N} = N! \chi_N$. 
The deviations from  ideal boson characteristics are incorporated
in the normalization term $\chi_N$ obtained using ${\langle N  | N \rangle}$=1 as
\cite{Combescot2011,romb,Law,woot} 
\bea
\label{norm}
\chi_N^{B} &=& N! \sum_{1 \le j_1 \le j_2 \dots \le j_N}^S \; \; 
\prod_{k=1}^N  \lambda_{j_k} , \\ \label{norm2}
\chi_N^{F} &=& N! \sum_{1 < j_1 < j_2 \dots < j_N}^S \; \; 
\prod_{k=1}^N  \lambda_{j_k} ,
\eea 
where $\chi_N^{B}$=$\chi_N^{F}$=1 for ideal bosons at all $N$, and $\chi_N^F$=0
when the number of bi-fermions, $N$, exceeds  the number of
available fermionic single-particle states, $S$.
For bi-fermion bosons, 
$\chi_N^{F}$ can be interpreted combinatorially as the
probability  associated with $N$ entities
yielding different outcomes, when a property $j$ ($ 1 \le  j   \le  S$) 
is assigned to each entity. There are  however
differences between the two species as multiple occupation of modes are forbidden
in bi-fermions unlike in  the case of bi-bosons
which are diverse in terms of the occupation profile of the Schmidt modes.
In general, it is difficult to compute exactly the  normalization factor 
for both  bi-fermions and bi-bosons.

\subsection {Upper and a lower bound to the normalization 
ratio}\label{bound}

A  simple inequality involving the upper and a lower bound to the normalization 
ratio, which yields a measure of departure from ideal boson properties,
 was obtained as \cite{woot} 
\be 
1- P\cdot N \le \frac{\chi_{N+1}}{\chi_N} \le 1-P,
\label{ineq1}
 \ee
 where the lower bound decreases monotonically with $N$,
and vanishes at $P$ = $\frac{1}{N}$. The corresponding
 uniform state $\vec \Lambda^U$ arises from a finite number 
($\frac{1}{P}$) of Schmidt modes, with ${\chi_{\frac{1}{P}+1}}$ = 0.
The normalization ratio is minimized by a uniform distribution $\vec \Lambda^U$.
The state associated with the $N$-independent upper bound in Eq. \ref{ineq1} remains
unsaturated as the  real, saturable upper bound is smaller than $1-P$. 
The bound $1-P$ provides saturable form for the corresponding state
at $N=1$. By determining the Schmidt coefficients of those states that extremize the normalization ratio, a quantitative indicator for bosonic behavior can be determined
 in terms of the purity $P$ and the number of composites in the same state $N$ \cite{woot,Tichy12}, 
\be 
1- P\cdot N \le \frac{\chi_{N+1}}{\chi_N} \le 1- \frac{P N}{1+(N-1) \sqrt{P}} 
\label{ineq2} .
\ee
These bounds will be useful in estimating physical quantities such as scattering rates,
and other processes in which the number of cobosons $N$ and 
single-fermion states $S$  remain large.

\subsection {Number-operator for composite bosons}\label{numb}

The physical interpretation of the mean number operator
$\hat N$ defined as
\be
\label{noper}
\hat N =  \hat c^\dag \hat c  =: \sum_{j,k=1}^S \sqrt{\lambda_j \;\lambda_k }\;\hat d^\dagger_j \;\hat d_k,
\ee
 is only unambigiuous when the constituents are highly entangled. However, with  increasing deviations from 
the ideal commutation relation, this expectation value operator yields a boson number that is less than the  total number of bi-fermions 
provided by the number-conserved operator
\be 
\label{binum}
\hat n_{tot}= \sum_{j=1}^S \; \hat n_j .
\ee
While $\hat n_j$ measures the number of bi-fermions or bi-bosons in a single
mode, $j$, it is not influenced by the  bosonic quality or entanglement attributes 
of the  composite bosons. The operator, $\hat n_j$ is number-conserving as
the number of   bi-fermions is conserved
under all dynamical processes, which includes those that unbind the
constituents into freely existing form. 
The apparent loss in the boson number  which appears in the  mean number operator
$\hat N$,  can be attributed to transitions of non-ideal fermionic fragments to
orthogonal subspaces which accommodate non-ideal states
 orthogonal to all other states $\ket{M}$ with $M=0,\ldots,N$.
The expectation value of 
the number operator $\hat N$ yields the number of bi-fermions
that exist as correlated entities, which differs from the interpretation
of  $\hat n_{tot}$ in Eq. \ref{binum} which
obeys an invariance in the  boson number. 
In this regard, the term ``number" 
holds different meanings for the two operators,
$\hat N$ and $\hat n_{tot}$, with the former operator associated with
the total number of composite bosons which are entangled
or remain correlated. On the other hand, $\hat n_{tot}$
includes all  constituents of the coboson, independent
of their state of correlation or existence as free fermions.
Here  we employ $\hat N$ as a coboson number
operator that quantifies only the correlated electron-hole pairs,
and which is amenable to change with environmental conditions.
We also utilize this operator within the 
BCS wave function ansatz associated with a grand canonical ensemble  to analyze
the  scattering  of excitons examined in this study.

In material systems such as semiconductors, the  coboson operator $\hat N$ effectively differentiates
strongly bound bosonic excitons
from free electron-hole pairs.
 With increase in fermion densities, the actual  number of bi-fermion
 pairs that can be treated as ideal bosons \cite{dis} decreases,
this is  reflected in a decreased expectation of $\hat N$
associated with lower  normalization ratios
 of the quantum state of $N$ composites. 
The difference between $\hat N$ and $\hat n_{tot}$  can be taken
as a measure of the non-ideal nature of cobosons.
For bi-fermions, we can set $\hat n_j = \hat d^\dag \hat d$
as each mode can only be occupied by at most a single bi-fermion.
The scenario is different in the case of 
bi-bosons as each mode $j$ can be occupied by several particles.
The expectation value of $\hat d^\dag \hat d$ yields $n_j^2$ instead of 
$n_j$. As a consequence, the expectation value of $\hat N$ for bi-boson composites can be larger than the actual number of bi-bosons, for which a physical interpretation is desirable. These differences highlight the challenges in treating bi-bosons
in the same footing as bi-fermion cobosons. We therefore  pay greater
attention to the scattering of bi-fermion condensates in this work, and consider
the bi-bosons on qualitative terms in Section \ref{sub}. 

\subsection {Formation of a fermionic fragment}\label{ferm}

The  process in which a  particle is removed from a coboson condensate
  occurs in a  total Hilbert space that is decomposed into two
 orthogonal subspaces. One subspace holds  the boson condensate while the other
is occupied by the orthogonal
 fragment species. The Fock-space with $N$ bi-fermions is  made up 
of an $N$-coboson-state and a fermionic non-ideal state that is
orthogonal to all coboson states.
The action of the creation operator, 
$\hat c^\dagger$ (Eq.\ref{boper}) on a $N$-composite bosons state can be derived as 
\be
\label{CreationAction}
\hat c^\dagger \ket{N} = \frac{\hat c^\dagger}{\sqrt{ N! \chi_N}}\ket{\psi_N} = \frac{1}{\sqrt{ N! \chi_N}}\ket{\psi_{N+1}} 
= \alpha_{N+1} \sqrt{N+1}\ket{N+1}
\ee
where $\alpha_N=\sqrt{\chi_{N}/\chi_{N-1}}$. 
The $\ket{N}$ state constitutes a subset of the entire Hilbert space 
associated with the constituent particles,  thus the action of $\hat c$ on $\ket{N}$ appears as 
\be
\label{cKetN}
\hat c \ket{N} = A_N \ket{N-1} + \ket{\varepsilon_N}
\ee
where $\ket{\varepsilon_N}$ denotes the fragment state  that is 
orthogonal to $\ket{N-1}$.
 The constant $A_N$ is obtained using  \ref{CreationAction} as
\be
\label{anc}
A_N= \bra{N-1}\hat c \ket{N} = \alpha_N \sqrt{N}.
\ee
The  state $\ket{\varepsilon_N}$ in Eq. \ref{cKetN} 
is orthogonal not only to the state $\ket{N-1}$, but also to any state $\ket{M}$ with $M=0,\ldots,N$ \cite{Law},
hence $\braket{M}{\varepsilon_N} = 0$ for $M = 0,\ldots,N$.
The correction factor, $\braket{\varepsilon_N}{\varepsilon_N}$ has been obtained as  \cite{Law,Combescot2011}
\be
\braket{\varepsilon_N}{\varepsilon_N} =1 - \frac{\chi_{N+1}}{\chi_N}
 - N \left( \frac{\chi_{N}}{\chi_{N-1}} 
- \frac{\chi_{N+1}}{\chi_N} \right).
\ee
For ideal bosons, $\braket{\varepsilon_N}{\varepsilon_N} \to 0$, and in the case
of bi-fermion cobosons such as excitons, the increased densities of  electron-holes pairs 
will result in a  higher correction factor, as the 
ratio, $\alpha_N$  is strictly non-increasing with $N$ \cite{woot}.

\section{Fluctuation to the mean number, $\langle \hat N \rangle$ of bi-fermions}\label{fluc}

In the context of the scattering process to be examined in this work,
the fluctuations in the mean number of correlated coboson constituents, $\langle \hat N \rangle$ present
as an important factor which quantifies  changes
that may occur  during dynamical interactions with external entities such as phonons.
While  the fluctuations measures changes in 
the correlated coboson constituents, it is possible that
the total number of fermion pairs (as measured by $\hat n_{tot}$ in Eq. \ref{binum})
 may be altered due to recombination effects that result in phonon emission. 
In this work, we assume that such recombination effects are minimal, and
focus on the influence of the normalization ratio on $\langle \hat N \rangle$  and fluctuations
associated with the number of correlated bi-fermion pair systems.

In an earlier work examining the commutation relations involving cobosons \cite{compra1,compra2},
a  relation was obtained as 
\be
\label{cdaggercket}
\hat c^\dagger \hat c \ket{\psi_N} = \ket{\psi_{N}} + \frac{N-1}{N+1} \hat c \ket{\psi_{N+1}}
\ee
Eq. \ref{cdaggercket} is  useful both in the calculation of
the  effective mean number, $\langle \hat N \rangle$ of bi-fermions
and in seeking extensions of the trilinear commutation
relations \cite{para1,para2,para3} to coboson
systems.
Using Eq.~\ref{cdaggercket} we obtain 
\bea
\label{n1}
\langle \hat N \rangle &=& \bra{N} \hat c^\dagger \hat c \ket{N} = \frac{\bra{\psi_N} \hat c^\dagger \hat c \ket{\psi_N}}{\braket{\psi_N}{\psi_N}} =
1+(N-1) \frac{\chi_{N+1}}{\chi_N} 
\\
\nonumber
\langle \hat N^2 \rangle &=& \frac{\bra{\psi_N} \hat c^\dagger \hat c \hat c^\dagger \hat c \ket{\psi_N}}{\braket{\psi_N}{\psi_N}} =
1+ \left( \frac{(N-1)^2}{N+1}+2N-2 \right) \frac{\chi_{N+1}}{\chi_N} + \frac{N(N-1)^2}{N+1} \frac{\chi_{N+2}}{\chi_N}
\\
\label{n2}
\eea
where $\langle \hat A \rangle =\bra{N} \hat A\ket{N}=\bra{\psi_N} \hat A \ket{\psi_N}/\braket{\psi_N}{\psi_N}$ 
is the mean value of the operator $\hat A$ and $\hat N = \hat c^\dagger \hat c$ is considered the cobosons number operator. 
We reiterate, as discussed in Section \ref{numb}, that $\hat N$ quantifies the number of excitons (or correlated bi-fermions)
and is not  inclusive of  the free electron-hole pairs which result from the scattering process to be 
considered shortly. 

For moderate
values of the  purities, $P = \frac{\gamma}{N}$ where $\gamma < 1$, we
obtain using Eqs.\ref{n1} and \ref{n2}, the 
fluctuation in the mean number,
$\langle \hat{N}\rangle$ as follows
\bea
\label{flucexp}
\left[ \frac{\langle \hat{N^2} \rangle-\langle
 \hat{N}\rangle^2}{\langle \hat{N}\rangle^2}\right ]_{\frac{\chi_{N+1}}{\chi_N}=1-P} &=& \quad
 \frac{\gamma  (N-1)^2 (N-\gamma )}{(N+1) \left(\gamma +N^2-\gamma  N\right)^2}
\\
\left[ \frac{\langle \hat{N^2} \rangle-\langle
 \hat{N}\rangle^2}{\langle \hat{N}\rangle^2}\right ]_{\frac{\chi_{N+1}}{\chi_N}=1-N P} &=& \quad
\frac{\gamma  (N-1)^2 (\gamma +(\gamma -1) N)}{(N+1) \left((\gamma -1) N^2\right)^2-\gamma}
\eea
with the fluctuations vanishing in the limit $P \rightarrow 0$, 
and increasing gradually with $P$. The expression for $\langle \hat{N}\rangle$
at the tighter upper bound (see Eq. \ref{ineq2}) is lengthy, and therefore
we do not include its form here.
While the bounds on $\chi_N$ also bound $\hat N$, this property does not extend to the
case of the fluctuations in the mean number, $\langle \hat{N}\rangle$.
The (normalised) second order correlator~$g_N^{(2)}$ 
characterizes the probability of detecting of particles
at times $t$ and $t+\tau$ \cite{glaub,lauss} 
\be
 \label{g2}
g_N^{(2)}(\tau)={\langle\ud{\hat c}(t)\ud{\hat c}(t+\tau)\hat c(t+\tau)\hat c(t)
\rangle\over\langle\hat N(t)\rangle\langle\hat N(t+\tau)\rangle}
\ee
Eq. \ref{g2} can be interpreted as a measure of 
correlations between $N$ cobosons, with exclusion of all free fermion constituents,
and takes into account 
 the time-dependence of creation and annihilation operators.
$g_N^{(2)}(\tau)$ is not directly interpretable in terms of the 
normalization ratio, $\frac{\chi_{N+1}}{\chi_N}$ and purity, $P$
due to the time independence of the latter quantities.
It is therefore appropriate to consider the second order correlation function at zero time delay , $g_N^{(2)}(0)$
which  provides information on the underlying statistical features, such
as the Poissonian case ($g_2(0) = 1$) in coherent systems involving
a large number of Fock states. 
$g_N^{(2)}(0)$
is  a useful indicator
of the bosonic quality  and may be used
to monitor rate changes during scattering processes involving cobosons.
$g_N^{(2)}(0)$ is rewritten using Eq. \ref{g2} as
\be
\label{g2r}
g_N^{(2)}(0) = \frac{\langle\hat c^\dagger \hat c^\dagger \hat c \hat c \rangle}{\langle \hat c^\dagger \hat c \rangle^2} 
\ee
The full derivation of $g_N^{(2)}(0)$ and analysis of its upper and lower bounds
will be considered elsewhere, however
we will refer to its utility in connection with the  
BCS variational ansatz in Section \ref{ansatz}.

\section{Scattering of  composite exciton condensates due to lattice vibrations}\label{pho}

\subsection{The BCS variational ansatz}\label{ansatz}
The typical  exciton creation operator with the center-of-mass
momentum $K$ and an internal motion associated with the $1s$ state can be written as  \cite{thil1,thil2a,thil2b,taka} 
\begin{eqnarray}
\label{exfun}
C_K^{\dag}= \sum _{k_{\rm e}, k_{\rm h}}
\; \delta_{K,k_{\rm e} + k_{\rm h}} \; \;  
\phi_{1s}(\alpha_{\rm e} {k}_{\rm h}
- \alpha_{\rm h} {k_{\rm e}})  a_{{k}_{\rm e}}^\dagger \; 
 h_{k_{\rm h}}^{\dagger} 
\end{eqnarray}
where the spin parameters  have been dropped  for simplicity
and $\alpha_{{\rm e}} (\alpha_{{\rm h}})$ = $\frac{m_e}{M}$($\frac{m_h}{M}$),
where $m_e$ ($m_h$) is the electron (hole) mass and $M$ is the total mass of the carriers.
The electron (hole) wavevectors ${{k}_{{\rm e}}}$ (${k_{{\rm h}}}$) in Eq. \ref{exfun}
spans the Brillouin zone in the momentum space.
 $ a_{k}^{\dagger}$ and $ h_{k}^{\dagger}$ denote the respective
electron and hole  creation operators, which are linked as 
\be
 h_{k}^{\dagger} = a_{-k}
\ee
In Eq. \ref{exfun}, $\phi_{1s}(\alpha_{{\rm e}}{k_{\rm h}}- \alpha_{\rm h}{k_{\rm e}})$ 
denotes the $1s$ wavefunction of a hydrogen type system, which  depends 
on the relative electron-hole separation in  real space.
The excitonic wavefunction can  be written as
be written as
\be
\label{exs}
\ket{\Phi_{ex}} = C_K^{\dag} \ket{0}
\ee
where the vacuum state $\ket{0}$ denotes a completely filled valence band, and an empty conduction band.

A mean-field description of the exciton condensate,
analogous to the Bardeen-Cooper-Schrieffer (BCS) form \cite{lit}
is suitable to model   a system of interacting fermions \cite{keld,comnoz1,comnoz2}.
The wavefunction of the composite condensate of
bi-fermion pairs with 
zero center-of-mass momentum  appears in a normalized form \cite{keld,comnoz1,comnoz2}
\begin{equation}
\ket{\Phi_{\text{BCS}}}= \prod_{k}\left[ u(k)+ v(k) a_{k}^{\dagger}h_{-k}^{\dagger} \right]\ket{0},
\label{bcs}     
\end{equation}  
where the coefficients, $u(k), v(k)$ satisfy the normalization condition,
 $u^2(k)+ v^2(k)$ = 1.  A small  ratio $\frac{v(k)}{u(k)}$ = $\phi_k$ $\ll 1$ applies
at the low-density range of the bi-fermion system at which
$u(k) \approx u(0)$ = 1 for all $k$. 
Eq. \ref{bcs} represents a state in which the constituents of the bi-fermion
pair ($a_{k}^{\dagger},h_{-k}^{\dagger}$) are either both present or absent,
hence the species  remain correlated for the lifetime  of the bi-fermion complex. 
The ground state becomes
separable only if either $u(k)$ = 1 or $v(k)$ = 1 for all $k$, 
however, the values of $|v(k)|^2$ for which
the correlated electron-hole pair
system retains its excitonic features
is not  apparent in Eq. \ref{bcs}.
This state changes from a system of 
excitonic boson  gas  to that of a  
two-component plasma present at  high fermion densities.
Such a change is dependent on  system
parameters such as the size of confinement, 
exciton Bohr radius, and density of fermion pairs.
The number of coboson particles is therefore not
fixed for the state in Eq. \ref{bcs}.
In the case of the dilute bi-fermion condensates,
$\sum_k |v(k)|^2$ = $\sum_k \frac{\phi_k^2}{(1+\phi_k^2)} \approx N (\frac{a_B}{L})^3$
where $a_B$ is the exciton bohr radius,  $L$ is the confinement length
and $\phi_{1s}(k)$ is the wavefunction in a three-dimensional momentum space 
 given by $\phi_{1s}(k) = \sqrt{64 \pi a_B^2}/
{[1+ (k a_B)^2]^2}$. For the reduced units, $a_B$=$L$=1, $\sum_k |v(k)|^2 = N_t$,  the total number of 
bi-fermions pairs in the ground state.

The overlap term, $\mathcal{O} = |u^*(k)\;v(k)|$ deserves special mention as
it is determined by the coherence between the bi-fermion pairs. This term
assumes a significance role in electronic properties of 
the condensate, and it will be shown to influence the  scattering properties due
to lattice vibrations (Section \ref{phos}), and the dynamics of growth 
 of the bi-fermion condensate.
The  negativity  $\mathcal{N}(k) = u(k)\;v(k)$ was proposed \cite{bran} as 
an entanglement measure of the interacting charge carriers
using
\be
\label{bcsne}
E_N(BCS)= \sum_k \; \log[\mathcal{N}(k)] = \sum_k \; \log[u(k)\;v(k)]
\ee
The negativity \cite{nega}  is equivalent to another well known  entanglement
 measure known as the concurrence, $\mathcal{C}$.
The concurrence measure $\mathcal{C}$ for the 
qubit state ($u|0 0 \rangle + v|1 1 \rangle$) appear as $2 u v$.
The two entanglement measures  ($\mathcal{N}$, $\mathcal{C}$)
may be treated as thermodynamical attributes
of the BCS wave function ansatz  associated with a grand canonical ensemble of a fixed chemical potential.
Both measures can be compared to
the ratio, $\frac{\chi_{N+1}}{\chi_N}$ which quantifies the entangled state
of $N$ coboson state.  The maximally entangled state is described
by $\mathcal{N}$=1 correlates with  the ideal boson state,
$\frac{\chi_{N+1}}{\chi_N}$=1.

In a  dilute system of bi-fermion pairs, the relation $\sum_k |u^*(k)\;v(k)|^2 \approx  \sum_k |v(k)|^2$
 can be employed to estimate the 
overlap term, $\mathcal{O}^2= \sum_k |u^*(k)\;v(k)|^2$. 
Using the effective mean number, $\langle \hat N \rangle$ 
of bi-fermions in  Eq.~\ref{n1} we obtain
\be
\label{fact3}
\sum_k |u^*(k)\;v(k)|^2 = \bra{N} \hat c^\dagger \hat c \ket{N} = N {\alpha^2}_{N+1}= N \frac{\chi_{N+1}}{\chi_N}
\ee
which is applicable in systems of low purity, $P$.
 Eq. \ref{fact3} can be understood by noting that 
the coherence between the bi-fermions is diminished
as a result of the addition of the $(N + 1)$st coboson  due to 
the Pauli principle occurring with the likelihood of 
$1 - \frac{\chi_{N+1}}{\chi_N}$. The system of 
bi-fermions will have enhanced $\mathcal{O}^2$
for entangled fermions  where there is no competition 
for single-fermion states due to the  Pauli principle. 
 The overlap term $\mathcal{O}^2$
is expectedly maximized in a system with purity, $P$ = 0 (see Eq. \ref{ineq1}).
It is to be noted that Eq. \ref{fact3}  is based on the assumption
that the system of electron-hole pairs is dilute and  highly
entangled (with low $P$ values). This allows the results from the normalization factor of
 N-identical composite boson state to be linked to parameters of the BCS wave function ansatz
as shown in Eq. \ref{fact3}.

The overlap term $\mathcal{O}^2$ can also  be estimated
using the second order correlation function at zero time delay, $g_N^{(2)}(0)$
given in Eq. \ref{g2r}. An analytical  form for $\mathcal{O}^2$ can thus 
be obtained by noting
a simple form of the bosonic quality term \cite{lauss} obtained using 
 $ g_N^{(2)}(0)={\gamma_{N-1}^2\gamma_N^2\over N^2}$. 
In the case of  excitons with bohr radius $a_\mathrm{B}$ 
placed in  quantum dots of size $L$, with  small values of $\frac{a_\mathrm{B}}{L}$
and  $N \ll \frac{L}{a_\mathrm{B}}$, $\gamma_N =
 \sqrt{N}\sqrt{1- 2(N-1)(\frac{a_\mathrm{B}}{L})^2}$. Using $g_N^{(2)}(0)$ to estimate $\mathcal{O}^2$, we obtain the
approximate relation
\be
\label{fact3b} 
\sum_k |u^*(k)\;v(k)|^2 = {\alpha^2}_{N+1} \approx N \left(1- \frac{2 \;N \; a^2_\mathrm{B}}{L^2} \right),
\ee
which is applicable to a system of $N$ bosons in quantum dots with small $\frac{a_\mathrm{B}}{L}$.
The role of a similar term in an earlier work on the scattering of composite bosons
has been discussed in Ref.\cite{imag}.
Using Eq. \ref{fact3b}, we note that at larger $\frac{a_B}{L}$ values, increased confinement effects
yields diminished number  of correlated electron-hole pairs
due to the Pauli exclusion principle. An increase in the fermionic fragment size
coupled with a Mott-like transition occurs at higher densities,
and  results in the formation of
 an electron-hole plasma state.
Hence  increased deviations from the ideal boson characteristics 
due to a decrease in quantum dot size gives rise to a reduced 
coherence features due to lower values of $\mathcal{O}^2$.

\subsection{Rate of Scattering of  composite exciton condensates}\label{phos}

\begin{figure}
  \includegraphics[width=9cm]{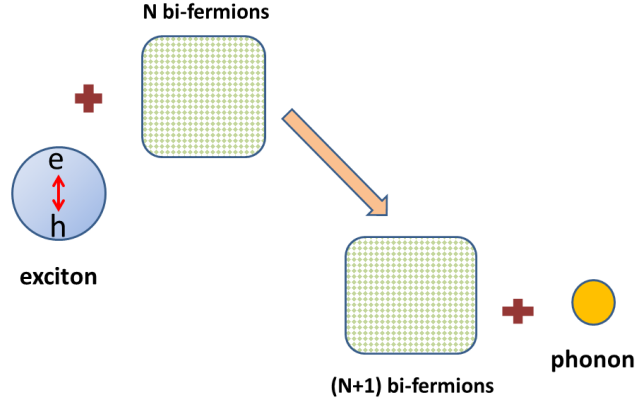}
\caption{Schematics of a 
channel in which an initial state  consisting  of  an
exciton and a composite condensate of $N$
bi-fermion pairs is scattered to a final state of  $N+1$ bi-fermions
pairs, with emission of a phonon.}
\label{scat}       
\end{figure}

We consider a process in which  an initial state   of an
exciton and a composite condensate of $N$
bi-fermion pairs gets scattered to a final state of  $N+1$ bi-fermions
pairs, with emission of a  phonon \cite{imag}.  The schematics of the 
channel is shown in Fig \ref{scat}. The momentum remains conserved
when the  exciton  plus condensate (C($0,N$)) system is scattered
to a final state of $(N+1)$ bi-fermion pairs, 
(C(0,N+1)), with creation of phonon with wavevector, $K-K'$ as follows
\be
{\rm Ex(K)+ C(0,N)} \rightarrow {\rm C(K',N+1) + phonon (K-K')}
\ee
The energy  of  the emitted phonon (with momentum $K-K'$)
is derived from the energy released
when the exciton coalesces with the condensate of bi-fermions. The final
state of bi-fermion condensate  acquires a net momentum of $K'$.
The composite exciton  Hamiltonian  in contact with a phonon reservoir  reads \cite{thil1}
\bea
\label{exHim}
\mathcal{H}_{T}&=& \mathcal{H}_{ex} + \mathcal{H}_{p}
+\mathcal{H}_{ep} ,\\
\label{phon}
\mathcal{H}_{p}&=&\sum_{ q} \hbar \omega({ q})b_{ q}^\dagger b_{ q} ,
\\
\nonumber
\mathcal{H}_{ep}&=& N^{-1/2}\sum_{ k,q}(\chi_e({ q})
a_{k+q}^\dagger a_{k}+\chi_h({ q})
h_{k+q}^\dagger h_{k}) (b_{- q}^\dagger +b_{ q}) ,
\label{exphon}
 \eea
where the exciton Hamiltonian $\mathcal{H}_{ex}$ is given by
$\sum_{ k} E_0({ k})C^{\dag}_k C_k$
and $E_0({ k})$ is the energy of the exciton in the absence of lattice fluctuations.
$\hat{H}_{p}$ denotes the
phonon energies and $b_{ q}^\dagger (b_{ q})$ is the creation (annihilation) phonon
operator with  frequency $\omega({ q})$ and  wavevector ${q}$. 
The exciton-phonon interaction operator, $\mathcal{H}_{ep}$ involves
the respective electron-phonon and hole-phonon 
coupling functions, $\chi_e$ and $\chi_h$.

The initial state consisting of an exciton and a composite condensate of 
bi-fermion pairs with zero center-of-mass momentum appear as \cite{comnoz1,comnoz2,keld,imag}
\begin{equation}
\ket{\Phi_i}= C_K^{\dag} \prod_{k}\left[ u(k)+v(k) a_{k}^{\dagger}h_{-k}^{\dagger} \right]\ket{0}
\label{istate}     
\end{equation}  
where the exciton possesses a center-of-mass momentum $K$ and an internal motion
described by the wave function $\phi_{1s}$ which appears in Eq. \ref{exfun}.
The final scattered state  becomes
\be
\ket{\Phi_f}= \prod_{k}\left[ u'(k)+v'(k) a_{k}^{\dagger}h_{-k}^{\dagger} \right]\ket{0},
\label{fstate}     
\ee
where $\sum_k |v'(k)|^2 \approx 1+N$, so that
$N+1$  bi-fermion pairs are present  in the final state. 
In the limit of  very large $N$ 
and low density condensates, $\frac{u(k)}{u(k')} \approx 1$
and $\frac{v(k)}{v(k')} \approx 1$, irrespective of the value of $k$.

The rate of scattering ($R_s$) of the process shown in Fig.\ref{scat}
can be obtained using the Fermi Golden Rule \cite{imag,thil2a,thil2b,taka}, assuming 
a large exciton momentum $K$. This ensures that there is 
no backflow of information from the reservoir due to  short memory bath times,
which allows the use of   the Born-Markov approximation 
\be
\label{scateq}
R_s=\frac{2 \pi N}{\hbar} \; \sum_{q} |\chi_e( 2 q)
\phi_{1s}(q)+\chi_e( 2 q) \phi_{1s}(- q)|^2 \; {\alpha^2}_{N+1} \; \delta(\hbar \omega_i-\hbar \omega_f-\hbar \omega_o)\\
\ee
where  $\alpha^2_{N+1}$ (using Eq. \ref{fact3}) quantifies
the effective probability of increasing the number of bi-fermion pairs
from a size of $N$ to $N+1$. 
$\hbar \omega_i$ ($\hbar \omega_f$) denotes the energy of the initial (final)
energy of the scattered system, and  $\hbar \omega_o$
is the energy of the  emitted phonon. In general, 
 Eq. \ref{scateq} is applicable to  small bosonic deviations which appear 
at low  bosonic densities, with simplification also introduced by neglecting
the $k$-dependence of the excitonic wave function from coherence terms such
as  $|u^*(k)\;v(k)|^2$. 
The upper bound for ${\alpha^2}_{N+1}$ 
(see Eq. \ref{ineq2}), indicates that the rate of scattering $R_s$ 
decreases with increase in purity $P$, in agreement with decreased probabilities
 of charge carriers  relaxing to  unoccupied states  due to the Pauli exclusion principle.

Based on the decrease of $\frac{\chi_{N+1}}{\chi_N}$
with $N$, we can conclude that the rate $R_s$ decreases with
increase in $N$  for bi-fermion cobosons.
The absence of phase-space of charge carriers, particularly near the Fermi level,
results in an inhibition of stimulated scattering processes
when coherence between the bi-fermion pair states is decreased.
There is the possibility that an uncorrelated electron-hole pair may bind to form
an exciton, with emission of phonons, however this process is less likely to 
occur in bi-fermion condensates of high $P$ values. 
As observed in an earlier work \cite{imag}, the  spontaneous and stimulated scattering
rates  decrease at larger densities at which greater deviations  from ideal bosonic behavior occur (at increased
values of $P$). In the limit of an an electron-hole plasma state, $\alpha^2_{N+1} \rightarrow$ 0,
and an absence of  stimulated emission is predicted.  

The appearance of the 
normalization ratio, $\alpha^2_{N+1}$ in Eq. \ref{scateq} is the main result of this work.
This ratio  captures the role of the Pauli exclusion principle at the point when 
there is competition for single-fermion states.
In an initial state, the $N$ bi-fermions could
occupy the modes $j_1 \dots j_N$, and the incoming
$N + 1$st coboson may need to be 
accommodated among the remaining $S- N$ unoccupied Schmidt
modes. The effective probability that the incoming bi-fermion 
occupies an initially  unoccupied Schmidt mode is evaluated
by adding all coefficients  associated with 
 the unoccupied mode configurations which is 
given by $\sum_{m \in j_{N+1},...,j_S} \; \lambda_m$.
This process has to be repeated for each configuration of
 $j_1,...,j_N$ to yield the final 
probability to add an N +1st coboson to an N-coboson
which is given by the 
normalization ratio, $\alpha^2_{N+1}$.
A redistribution
among the bi-fermion Schmidt modes may occur as a result of  scattering processes,
including those with no phonon emission, and the normalization
ratio may be affected by the outgoing phonon energies.
Such possibilities   need greater examination in future works.

Accurate values of $\alpha^2_{N+1}$ are  generally not easily computable for
two-fermion wavefunctions and large number $N$ of bi-fermion pair systems,
however the bounds obtained in Ref. \cite{tichy2013b} do resolve the
computational demands associated with  large boson systems.
An alternative measure that can be used  to assess the scattering process involves incorporation of
the fluctuations in the mean number,
$\langle \hat{N}\rangle$ (Eq.\ref{flucexp}) in the rate expression, $R_s$ (Eq.\ref{scateq}).
The scattering process
is optimized when fluctuations in the exciton number vanish in the limit $P \rightarrow 0$, due to the availability of a maximum number of
ideal bosonic excitons  for interaction with the phonons.
The qualitative predictions here may be tested following the experimental work of 
Mondal et. al. \cite{mond} who investigated 
the dynamics of  state-filling dynamics in self-assembled 
InAs/GaAs quantum dots (QDs) using picosecond excitation-correlation (EC) spectroscopy.
The  action of  the Pauli exclusion principle appeared visible in the photoluminescence results \cite{mond}. 
Future experimental works may wish to examine the  controlled scattering of 
excitons which occupy specific Schmidt modes, 
and the subsequent emission of phonons with a desired range of
 energies.

The strong relationship between quantum entanglement of the constituents
of boson systems and their bosonic quality therefore play an important 
role in the scattering process depicted in Fig. \ref{scat}, and as seen in the rate
$R_s$ of Eq. \ref{scateq}. The usefulness of the normalization 
term may be studied in scattering processes involving other generalized composite models, such as bi-bosons which
are made up of smaller boson fragments \cite{tichy2013a}.
The scattering dynamics
which occurs in the case of bi-boson systems will be considered in  Sec. \ref{sub}

\subsection{Application to the dynamics of singlet and triplet excitons}
In strategic polymer  materials, the dynamics of 
exciton is determined by the kinetic 
transformation involving singlet and triplet excitonic states \cite{pol1,pol2}.
While singlet excitons are emissive and account for  electroluminescence
in conjugated polymers, triplet excitons remain non-emissive and these differences in optical
properties give rise to a range of electroluminescence efficiencies in polymers.
It is therefore worthwhile to provide brief mention of
 the extension  of the scattering rate in
Eq.\ref{scateq} to excitons which can form in the  singlet or triplet state, depending
on the  spin angular momentum. It is known that four spin eigenstates can result
from  the electron-hole quasi-particle based on the spin 
angular momentum operator ${\mathcal S}$, and  its
$z-$component, ${\mathcal S}_z$ as follows
as
\bea
\label{sp}
&& (1) \quad \quad
a_{\uparrow} \; h_{\uparrow} \\ \nonumber
&& (2) \quad \quad
\frac{1}{\sqrt{2}} (a_{\downarrow} \; h_{\uparrow} + h_{\downarrow} \; a_{\uparrow}) \\
 \nonumber
&& (3) \quad \quad
a_{\downarrow} \; h_{\downarrow} \\ \nonumber
&& (4) \quad \quad
\frac{1}{\sqrt{2}} (a_{\downarrow} \; h_{\uparrow} - h_{\downarrow} \; a_{\uparrow})
\eea
The first three symmetric eigenstates of the triplet exciton in Eq. \ref{sp} are associated with ${\mathcal S}$=1,
while the last anti-symmetric state of the singlet exciton  is  linked to  ${\mathcal S}$=0.
Due to the Pauli exclusion principle, the
triplet state is correlated with the  anti-symmetric spatial
wavefuntion, while the  singlet state is linked with the  symmetric spatial
wavefuntion. On this basis, differences in 
the  probabilities of occupation of  Schmidt modes of
singlet and triplet excitons are to be expected, with likely
variations in the scattering rates for the two types of excitons.
Important mechanisms such as
the scattering of the triplet exciton into the 
singlet exciton state via
 acoustic phonons, as well as the 
fission of a singlet exciton into two triplet excitons \cite{fission}
are similarly expected to be influenced by the dependence
of $\chi_{N+1}/\chi_N$ on ${\mathcal S}$.
A detailed examination of the exact dependence of the
normalization ratio $\chi_{N+1}/\chi_N$, which governs
the  scattering rate in Eq. \ref{scateq}, on the operator
${\mathcal S}$  will be considered in future works.

\section{Composite boson made of two bound bosonic constituents (bi-bosons)}\label{sub}

The multiple occupation of single constituents in a specific
mode for bi-bosons is not compromised due to 
Pauli-blocking as is evident in Eq. \ref{norm}.
The bi-boson operator, $d_{j}^{\dagger}:= a_{j}^{\dagger} b_{j}^{\dagger}$,
satisfies \cite{tichy2013a}
\be \label{nocc} 
[d_{j}^{\dagger},d_{k}^{\dagger}]=\delta_{j,k}(1+2 n_j) 
\ee
as well as
the over-normalization relation, $[d_{j}^{\dagger}]^N \; \ket{0} = N! \ket{N}_j$ \cite{tichy2013a}.
where $n_j$ denotes the number of bi-bosons in the $j$th mode. These relations highlight
the enhanced bunching tendencies of the two-boson composites 
as there can be multiple occupation
of  a single Schmidt mode. The
normalization ratio for bi-boson composites appear as \cite{tichy2013a}
\be
\label{ineqb}
P N+1  \le \frac{\chi_{N+1}}{\chi_N}  {\le} \sqrt P N +1 ,  
\ee
which may be compared to Eq. \ref{ineq2}
 for bi-fermion type bosons. The relation in Eq. \ref{ineqb} 
is not saturable, however a relation with tight bounds
is not in the simple form provided here.

In bi-boson cobosons, 
there are  two regimes associated with $N \sqrt{P} \ll 1$ and 
$N \lambda_1 \gg 1$ \cite{tichy2013a} where $\lambda_1$ is the 
largest Schmidt coefficient. In the latter regime, the 
Schmidt modes with magnitude $\lambda_1$ are favorably
populated resulting in  the characteristic super-bunching tendency of bi-bosons.
 As the number of composites $N$ is increased,
a Schmidt mode that is occupied by a boson is likely to attract
further occupation due to the $n_j$ dependency in Eq. \ref{nocc}.
The increase in the effective boson number
 in the favored macroscopically occupied  Schmidt modes occurs at the expense of bosons
distributed in other modes or present in adjacent ortho-complement subspaces.

The bunching attributes of bi-bosons has implications for scattering processes as
an incoming boson species that collides with the main coboson target
is likely to occupy the macroscopically occupied  Schmidt mode
resulting in an increased rate. In this regard,
the scattering may differ from that involving
 a composite boson system of bi-fermion pairs where
the scattering rate decreases with 
increase in $N$ bi-fermions.
For scattering of phonons of select energies, there may
be enhanced scattering of bi-bosons at conditions favorable to super-bunching (such as large $N$)
which can be deduced using Eqs.\ref{scateq} and \ref{ineqb}.

The   quantum droplet  formed from  electron-hole aggregates \cite{dropt}
promises as a suitable platform to test the
quantum mechanical features such as bunching attributes of bi-bosons, under given experimental 
conditions. For instance, two excitonic droplets may be deposited in spatially separated quantum wells,
and depending on the inter-well   tunneling strengths and intra-well bosonic interactions,
 the presence of superbosonic  features  under controlled  conditions may be probed. 
Likewise the enhanced scattering properties of bi-bosons  involving phonons 
in integrated circuits that are subject to lattice vibrations, could also be investigated 
in future experimental works.

\section{Conclusion and Outlook}\label{con}

In this paper, we have examined the influence of Pauli-exclusion of fermions
when composite bosons of bi-fermion pairs undergo
scattering due to interactions with phonons. 
The entanglement
 between the fermionic constituents explicitly enters in the 
scattering rate of the composites. Large entanglement ($P \ll 1/N$) is 
synonymous for ideal bosonic behavior, while smaller entanglement 
leads to phase-space-filling effects, with reduced scattering.
Composite bosons characterized by larger purities (with high densities of bosons), 
have decreased scattering due to the phase-space filling effect,
where there is  decreased probabilities
 of charge carriers  relaxing to  unoccupied states.
The demonstration of the dependence of the scattering
rate on the normalization ratio, $\alpha^2_{N+1}$  highlights the  usefulness of 
the derived scattering rate  in the investigation of generalized
bosonic systems with multiple condensates such as quantum droplets \cite{dropt}.

When the composite boson under consideration is made of 
 smaller  boson fragments such as in the case of 
bi-bosons,  the  scattering process is predicted to reveal
 features that are qualitatively different
from those involving bi-fermion cobosons.
In particular, due to 
super-bunching properties of  bosons occupying
 macroscopically occupied Schmidt modes, there may
be enhanced scattering  linked to specific modes.
The results of this work contributes to fundamental aspects
of quantum mechanical modeling of composite boson systems. This study   has potential application 
in  Bose-Einstein condensates in confined systems,
and in the control of inter-bosonic carrier-carrier
interactions in photovoltaic technologies that rely on 
the mechanism of  multiple exciton generation (MEG) \cite{noz,bear}.
The Pauli exclusion principle  is also expected  to dominate exciton dynamics
in  layered transition  metal dichalcogenides
with show rich excitonic features \cite{mak13,ding11,thil14}.
To this end, the exclusion principle may be  examined  in
layered semiconductor systems in future works.

\section{Acknowledgements}
The author gratefully acknowledges the support of the Julian Schwinger Foundation Grant,
JSF-12-06-0000 and  would like to thank Malte Tichy,  Alexander Bouvrie and
Keun Oh for useful correspondences regarding
specific properties of composite bosons systems, and the 
anonymous referee  for 
helpful comments.

\eject


\begin{thebibliography}{10}
\expandafter\ifx\csname url\endcsname\relax
  \def\url#1{\texttt{#1}}\fi
\expandafter\ifx\csname urlprefix\endcsname\relax\def\urlprefix{URL }\fi
\expandafter\ifx\csname href\endcsname\relax
  \def\href#1#2{#2} \def\path#1{#1}\fi

\bibitem{compra1}
M.~Combescot, F.~Dubin, M.~Dupertuis, Role of fermion exchanges in statistical
  signatures of composite bosons, Physical Review A 80~(1) (2009) 013612.

\bibitem{compra2}
M.~Combescot, O.~Betbeder-Matibet, F.~Dubin, Mixture of composite-boson
  molecules and the pauli exclusion principle, Physical Review A 76~(3) (2007)
  033601.

\bibitem{comrep}
M.~Combescot, O.~Betbeder-Matibet, F.~Dubin, The many-body physics of composite
  bosons, Physics Reports 463~(5) (2008) 215--320.

\bibitem{Combescot2011}
M.~Combescot, Commutator formalism for pairs correlated through schmidt
  decomposition as used in quantum information, EPL (Europhysics Letters)
  96~(6) (2011) 60002--60007.

\bibitem{romb}
S.~Rombouts, D.~Van~Neck, K.~Peirs, L.~Pollet, Maximum occupation number for
  composite boson states, Modern Physics Letters A 17~(29) (2002) 1899--1907.

\bibitem{Law}
C.~Law, Quantum entanglement as an interpretation of bosonic character in
  composite two-particle systems, Physical Review A 71~(3) (2005) 034306.

\bibitem{woot}
C.~Chudzicki, O.~Oke, W.~K. Wootters, Entanglement and composite bosons,
  Physical review letters 104~(7) (2010) 070402.

\bibitem{avancini}
S.~Avancini, J.~Marinelli, G.~Krein, Compositeness effects in the
  bose--einstein condensation, Journal of Physics A: Mathematical and General
  36~(34) (2003) 9045.

\bibitem{Sancho}
P.~Sancho, Compositeness effects, pauli's principle and entanglement, Journal
  of Physics A: Mathematical and General 39~(40) (2006) 12525.

\bibitem{Rama1}
R.~Ramanathan, P.~Kurzynski, T.~K. Chuan, M.~F. Santos, D.~Kaszlikowski,
  Criteria for two distinguishable fermions to form a boson, Physical Review A
  84~(3) (2011) 034304.

\bibitem{Rama2}
P.~Kurzy{\'n}ski, R.~Ramanathan, A.~Soeda, T.~K. Chuan, D.~Kaszlikowski,
  Particle addition and subtraction channels and the behavior of composite
  particles, New Journal of Physics 14~(9) (2012) 093047.

\bibitem{brou}
T.~Brougham, S.~M. Barnett, I.~Jex, Interference of composite bosons, Journal
  of Modern Optics 57~(7) (2010) 587--594.

\bibitem{Gavrilik}
A.~Gavrilik, Y.~A. Mishchenko, Entanglement in composite bosons realized by
  deformed oscillators, Physics Letters A 376~(19) (2012) 1596--1600.

\bibitem{Gav2}
A.~Gavrilik, Y.~A. Mishchenko, Energy dependence of the entanglement entropy of
  composite boson (quasiboson) systems, Journal of Physics A: Mathematical and
  Theoretical 46~(14) (2013) 145301.

\bibitem{Tichy12}
M.~C. Tichy, P.~A. Bouvrie, K.~M{\o}lmer, Bosonic behavior of entangled
  fermions, Physical Review A 86~(4) (2012) 042317.

\bibitem{thilzil}
A.~Thilagam, Binding energies of composite boson clusters using the szilard
  engine, arXiv preprint arXiv:1309.6493.

\bibitem{thilchem}
A.~Thilagam, Crossover from bosonic to fermionic features in composite boson
  systems, Journal of Mathematical Chemistry 51~(7) (2013) 1897--1913.

\bibitem{tichyprl}
M.~C. Tichy, P.~A. Bouvrie, K.~M{\o}lmer, Collective interference of composite
  two-fermion bosons, Physical review letters 109~(26) (2012) 260403.

\bibitem{tichy2013a}
M.~C. Tichy, P.~A. Bouvrie, K.~M{\o}lmer, Two-boson composites, Physical Review
  A 88~(6) (2013) 061602.

\bibitem{tichy2013b}
M.~C. Tichy, P.~A. Bouvrie, K.~M{\o}lmer, How bosonic is a pair of fermions?,
  Applied Physics B (2013) 1--12.

\bibitem{thil1}
I.-K. Oh, J.~Singh, A.~Thilagam, A.~Vengurlekar, Exciton formation assisted by
  lo phonons in quantum wells, Physical Review B 62~(3) (2000) 2045.

\bibitem{thil2a}
A.~Thilagam, Exciton-phonon interaction in fractional dimensional space,
  Physical Review B 56~(15) (1997) 9798.

\bibitem{thil2b}
A.~Thilagam, J.~Singh, Generation rate of 2d excitons in quantum wells, Journal
  of luminescence 55~(1) (1993) 11--16.

\bibitem{hana}
E.~Hanamura, H.~Haug, Condensation effects of excitons, Physics Reports 33~(4)
  (1977) 209--284.

\bibitem{taka}
T.~Takagahara, Localization and energy transfer of quasi-two-dimensional
  excitons in gaas-alas quantum-well heterostructures, Physical Review B
  31~(10) (1985) 6552.

\bibitem{ina}
T.~Inagaki, M.~Aihara, Many-body theory for luminescence spectra in
  high-density electron-hole systems, Physical Review B 65~(20) (2002) 205204.

\bibitem{chuchang}
H.~Chu, Y.~Chang, Theory of optical spectra of exciton condensates, Physical
  Review B 54~(7) (1996) 5020.

\bibitem{bar1}
J.~Bardeen, L.~N. Cooper, J.~R. Schrieffer, Theory of superconductivity,
  Physical Review 108~(5) (1957) 1175.

\bibitem{bar2}
J.~Bardeen, L.~N. Cooper, J.~R. Schrieffer, Microscopic theory of
  superconductivity, Physical Review 106~(1) (1957) 162--164.

\bibitem{koi}
Z.~Koinov, Secondary peak in the optical absorption spectra: A possible
  criterion for bose condensation of excitons, Physics Letters A 371~(4) (2007)
  322--326.

\bibitem{imag}
A.~Imamo{\u{g}}lu, Phase-space filling and stimulated scattering of composite
  bosons, Physical Review B 57~(8) (1998) R4195.

\bibitem{dropt}
A.~Almand-Hunter, H.~Li, S.~Cundiff, M.~Mootz, M.~Kira, S.~Koch, Quantum
  droplets of electrons and holes, Nature 506~(7489) (2014) 471--475.

\bibitem{schmi}
E.~Schmidt, Math. Annalen 63 (1906) 433.

\bibitem{grobe}
R.~Grobe, K.~Rzazewski, J.~Eberly, Measure of electron-electron correlation in
  atomic physics, Journal of Physics B: Atomic, Molecular and Optical Physics
  27~(16) (1994) L503.

\bibitem{dis}
M.~Combescot, X.~Leyronas, C.~Tanguy, On the n-exciton normalization factor,
  The European Physical Journal B-Condensed Matter and Complex Systems 31~(1)
  (2003) 17--24.

\bibitem{para1}
H.~Green, A generalized method of field quantization, Physical Review 90~(2)
  (1953) 270.

\bibitem{para2}
O.~Greenberg, A.~Messiah, Selection rules for parafields and the absence of
  para particles in nature, Physical Review 138~(5B) (1965) B1155.

\bibitem{para3}
A.~Thilagam, M.~A. Lohe, Coherent states and their time dependence in
  fractional dimensions, Journal of Physics A: Mathematical and Theoretical
  40~(35) (2007) 10915.

\bibitem{glaub}
R.~J. Glauber, The quantum theory of optical coherence, Physical Review 130~(6)
  (1963) 2529.

\bibitem{lauss}
F.~P. Laussy, M.~M. Glazov, A.~Kavokin, D.~M. Whittaker, G.~Malpuech,
  Statistics of excitons in quantum dots and their effect on the optical
  emission spectra of microcavities, Physical Review B 73~(11) (2006) 115343.

\bibitem{lit}
X.~Zhu, P.~Littlewood, M.~S. Hybertsen, T.~Rice, Exciton condensate in
  semiconductor quantum well structures, Physical review letters 74~(9) (1995)
  1633.

\bibitem{keld}
L.~Keldysh, A.~Kozlov, Collective properties of excitons in semiconductors,
  Sov. Phys. JETP 27 (1968) 521.

\bibitem{comnoz1}
C.~Comte, P.~Nozieres, Exciton bose condensation: the ground state of an
  electron-hole gas i. mean field description of a simplified model, Journal de
  Physique 43~(7) (1982) 1069--1081.

\bibitem{comnoz2}
P.~Nozieres, C.~Comte, Exciton bose condensation: the ground state of an
  electron-hole gas ii. spin states, screening and band structure effects,
  Journal de Physique 43~(7) (1982) 1083--1098.

\bibitem{bran}
F.~G. Brand{\~a}o, Entanglement and quantum order parameters, New Journal of
  Physics 7~(1) (2005) 254.

\bibitem{nega}
G.~Vidal, R.~F. Werner, Computable measure of entanglement, Physical Review A
  65~(3) (2002) 032314.

\bibitem{mond}
R.~Mondal, B.~Bansal, A.~Mandal, S.~Chakrabarti, B.~Pal, Pauli blocking
  dynamics in optically excited quantum dots: A picosecond
  excitation-correlation spectroscopic study, Physical Review B 87~(11) (2013)
  115317.

\bibitem{pol1}
Y.~Cao, I.~D. Parker, G.~Yu, C.~Zhang, A.~J. Heeger, Improved quantum
  efficiency for electroluminescence in semiconducting polymers, Nature
  397~(6718) (1999) 414--417.

\bibitem{pol2}
M.~Wohlgenannt, K.~Tandon, S.~Mazumdar, S.~Ramasesha, Z.~Vardeny, Formation
  cross-sections of singlet and triplet excitons in $\pi$-conjugated polymers,
  Nature 409~(6819) (2001) 494--497.

\bibitem{fission}
P.~M. Zimmerman, F.~Bell, D.~Casanova, M.~Head-Gordon, Mechanism for singlet
  fission in pentacene and tetracene: From single exciton to two triplets,
  Journal of the American Chemical Society 133~(49) (2011) 19944--19952.

\bibitem{noz}
A.~J. Nozik, Spectroscopy and hot electron relaxation dynamics in semiconductor
  quantum wells and quantum dots, Annual review of physical chemistry 52~(1)
  (2001) 193--231.

\bibitem{bear}
M.~C. Beard, R.~J. Ellingson, Multiple exciton generation in semiconductor
  nanocrystals: Toward efficient solar energy conversion, Laser \& Photonics
  Reviews 2~(5) (2008) 377--399.

\bibitem{mak13}
K.~F. Mak, K.~He, C.~Lee, G.~H. Lee, J.~Hone, T.~F. Heinz, J.~Shan, Tightly
  bound trions in monolayer mos2, Nature materials 12~(3) (2013) 207--211.

\bibitem{ding11}
Y.~Ding, Y.~Wang, J.~Ni, L.~Shi, S.~Shi, W.~Tang, First principles study of
  structural, vibrational and electronic properties of graphene-like mx2 (m=
  mo, nb, w, ta; x= s, se, te) monolayers, Physica B: Condensed Matter 406~(11)
  (2011) 2254--2260.

\bibitem{thil14}
A.~Thilagam, Exciton complexes in low dimensional transition metal
  dichalcogenides, Journal of Applied Physics 116~(5) (2014) 053523.

\end{thebibliography}
\end{document}